\documentclass[preprint,pra,showkeys]{revtex4}
\newcommand{\be}{\begin{equation}}
\newcommand{\ee}{\end{equation}}
\newcommand{\bea}{\begin{eqnarray}}
\newcommand{\eea}{\end{eqnarray}}

\newcommand{\nn}{\mbox{$\nonumber$}}

\newcommand{\w}{\mbox{$\omega$}}

\begin{document}

\title{Can Light signals travel faster-than-c in non-trivial vacua \newline in flat space-time?
Relativistic Causality II.}

\author{Heidi Fearn}

\affiliation{Department of Physics \\
California State University Fullerton \\
Fullerton CA 92834 USA. }


\email{hfearn@fullerton.edu }
\date{\today}

\begin{abstract}

In this paper we show that the Scharnhorst effect (Vacuum with
boundaries or a Casimir type vacuum) cannot be used to generate
signals showing measurable faster-than-c speeds. Furthermore, we
aim to show that the Scharnhorst effect would violate special
relativity, by allowing for a variable speed of light in vacuum,
unless one can specify a small invariant length scale. This invariant
length scale would be agreed upon by all
inertial observers. We hypothesize the approximate
scale of the invariant length.

\end{abstract}



\maketitle


\section{Introduction}

\noindent
This is the second paper on the theory of superluminal light propagation by the author,
the first paper considered material media (non-vacuum), \cite{heidi1,heidi2}.

There has been much interest in recent theory and
experiment on superluminal light propagation in media
\cite{sci,aep,eric,wang,kuz,mil,segev}. These articles appear not
to violate relativistic causality, by which we mean the front
velocity cannot travel faster--than--c. The front velocity is the
speed at which the very first extremely small vibrations of the
wave will occur \cite{bril}. The front of the wave is composed of
the highest uv frequency components (ie $\w \rightarrow \infty$).
Sommerfeld \cite{somm}, predicted that the front velocity should
travel at c. A brief history of events is given in Brillouin's
book \cite{bril}. Later Voigt \cite{bril}, gave a physical
explanation for Sommerfelds result, stating that the modern theory
of dispersion uses the assumption of point like electrons with a
finite mass. Inertia prevents the electrons from oscillating at
the very start of the wave. Electron oscillation needs time to
build; only after the wave has been in motion for some time can
the electrons react back on the wave. The very early oscillations
of the wave therefore pass though the dispersive medium as if
through vacuum. Also, electrons having a finite mass, cannot
oscillate at infinite frequencies without amassing infinite energy
in the process. We will restrict ourselves here to the vacuum, in flat space.
We will need to treat the case of the vacuum in highly curved space-time in
a later paper. For now we wish to consider the Scharnhorst effect.

\section{Scharnhorst Effect}

\noindent
The Scharnhorst effect relates to light propagating in the vacuum
between two parallel mirrors.
The vacuum modes are changed by the boundaries
(in much the same way as in the Casimir effect) and the light
experiences the vacuum as a dispersive birefringent medium.
The real part of the refractive index in a direction parallel to the
mirror surface is unity. The refractive index in a direction perpendicular
to the mirror surface is found to be less than unity.
These calculations are done using perturbation theory valid only
for small frequencies $\w << m$ where $m$ is the electron mass.\\

Scharnhorst \cite{scharn,scharn2}, derived a refractive index (perpendicular
to the mirror surface) for the vacuum based on low
frequencies and showed $n(0) <1$. (Note that $n(0)$ implies
the small frequency limit.) Combining this with the Kramers Kronig
relations written for\\
\noindent
Re $n(\w) -$ Re $n(\infty)$ (as opposed to
Re [$n(\w) -1$]), and setting $\w =0$, Barton and Scharnhorst
\cite{barsh} showed that either (real part) Re $n(\infty) < 1$,
which would imply
signals moving at faster--than--c speed, or (imaginary part)
Im $n(\w) <0$ for some frequency range, which would imply
that the vacuum could amplify the light signal for some range of frequency.
The paper \cite{barsh}, does not make a distinction between these two
choices. \\

Here I would like to point out that the dispersion relations used by
Barton and Scharnhorst \cite{barsh} and later by Scharnhorst alone
\cite{scharn2} are not complete, they are approximate as will be shown below.

\noindent
Scharnhorst has stated \cite{scharn2} that it is not possible to
derive the result for $n(\infty)$ directly because it would require
a non--perturbative calculation. Therefore, they have attempted to
derive $n(\infty)$ by using the dispersion relations.
Barton and Scharnhorst \cite{barsh,scharn2} set the imaginary part of $n(\infty)=0$
in their papers. We cannot be sure if that is true or not, which they also admit.\\

Following Barton and Scharnhorst \cite{scharn2} the real part of $n(\w)$ is given in their paper as;

\be
\mbox{Re } n(\w) = \mbox{ Re } n(\infty) + \frac{2}{\pi}
P \int_{0}^{\infty}  d \w ' \;\; \frac{ \w'\; \mbox{Im }n (\w')}{ \w'^2 - \w^2}
\label{nw}
\ee

where they have assumed that $ \mbox{ Im } n(\w' \rightarrow \infty )=0$. Interestingly,
Barton and Scharnhorst employ Titchmarch's 1948 theorem which does have the added implication that
$ n(\w) \rightarrow 1$ as $\w \rightarrow \infty $ \cite{heidi1, heidi2} to ensure
relativistic causality. There is a possibility that Titchmarsh's theorem has been incorrectly cited
in some texts and papers since the original theorem is spread out in his book \cite{titch} and covers
several sections. Also back in 1948 no one was considering faster than c signals, so some of the
implications of the theorem may have been overlooked.

\noindent
We point out here that the above dispersion relation has been simplified, and is of the form used
when a physical model for a material (like the Drude model) is used. We are not convinced that
such a physical model applies to the vacuum.
The most general form of the dispersion relation can be derived as follows;
if the square integrability condition on some function $G(\w)$ cannot be satisfied,
but the weaker condition that $G(\w)$ is bounded, $|G(\w)|^2 \leq K_0$, where $K_0$ is a finite
constant can be satisfied, then we may construct a new function
\be
H(\w) = \frac{ G(\w) - G(\w_0)}{\w - \w_0} \mbox{\hspace{1.0in} Im }\w_0 \geq 0
\ee
where $H(\w)$ is square integrable and has no poles in the upper half of the complex plane
$I_+$  and hence satisfies the dispersion relations.
\bea
H(\w) &=& \frac{P}{i\pi} \int_{-\infty}^{\infty} \frac{ H(\nu)}{\nu - \w} d\nu
\mbox{ \hspace{1.0in} real $\w$} \nn \\
G(\w) &=& G(\w_0) + \frac{\w -\w_0}{i \pi} P \int_{-\infty}^{\infty}
\frac{ G(\nu) - G(\w_0)}{\nu -\w_0} \frac{ d\nu}{ \nu - \w}
\eea
where $G(\w_0)$ is called the subtraction constant and $P$ indicates the principal
part of the integral. Taking the real part we find,
\be
\mbox{Re } G(\w) = \mbox{ Re } G(\w_0) + \frac{\w -\w_0}{\pi}
P \int_{-\infty}^{\infty} \mbox{Im } \left[
\frac{ G(\nu) - G(\w_0)}{\nu -\w_0} \right] \frac{ d\nu}{ \nu - \w}
\ee
\noindent
This is known as a dispersion relation for $G(\w)$ with one subtraction. Often,
subtractions will occur at $\w_0 = 0$ or $\w_0 = \infty$.
More than one subtraction is allowed. Using the $\w_0 = \infty$ subtraction
for $n(\w)$ we find,
\bea
\mbox{Re } n(\w) &=& \mbox{Re }n(\infty) + \lim_{\w_0 \rightarrow \infty}
\left[ \frac{ \w -\w_0}{\pi} P \int_{-\infty}^{\infty} \mbox{ Im } \left[
\frac{ n(\nu) - n(\w_0) }{\nu - \w_0} \right] \frac{ d\nu}{\nu - \w} \right]
\nn \\
&=& \mbox{Re }n(\infty) +
\frac{P}{\pi} \int_{-\infty}^{\infty} \mbox{ Im }
\left[ n(\nu) - n(\infty) \right] \frac{ d\nu}{\nu - \w} \nn \\
&=& \mbox{Re } n(\infty) + \frac{2P}{\pi} \int_0^{\infty}
\left[ \nu \mbox{ Im }n(\nu) - \w \mbox{ Im }n(\infty) \right]
\frac{d\nu}{\nu^2 -\w^2}
\label{lasteqn}
\eea
\noindent
where it has been assumed that Im $n(\nu)$ is odd in the last line. Do we know it to be odd?
This assumption comes from a physical model.
We are using $\nu$ here instead of $\w'$ if one wishes to compare with the Barton and Scharnhorst
equation [\ref{nw}].
This is the same as the result implied by Landau and Lifshitz
\cite{stat,ll}, but it only requires primitive causality. For relativistic
causality we also require that $n(\infty)=1$ so that the front velocity of
a signal travels at $c/1$ and not some arbitrary $c/n(\infty)$.\\

Some readers may consider these approximations are reasonably trivial and can be overlooked
but a further objection to the Scharnhorst effect is that it can violate special relativity (SR).
Several people I discussed this with at the conference CCFP'06 believe this to be true.
It is certainly worth further investigation.
Even if you cannot measure the Scharnhorst effect, to be discussed below,
the fact that it can exist even in principle is objectionable.\\

\noindent
In the following sections, we will shall assume the Scharnhorst effect is correctly derived, implying
that the dispersions relations are valid for the bounded vacuum, no matter what length is involved.
Usually, when dispersion relations are employed for a material medium, some length scale is implied
for the validity of the use of a refractive index to represent the material in ``bulk". The wavelength
of the light under observation must be much larger than the interatomic spacing for the refractive index
approximation to be valid. No such ``interatomic spacing" or equivalent length scale is immediately
obvious in the case of the vacuum.
There is no length scale specified in the original papers of Scharnhorst and Barton, and so we can
assume any scale we wish. The calculated change in the velocity of light (even if it cannot be measured
in practice) is still deeply disturbing since it appears to violate fundamental axioms of quantum
electrodynamics (QED). We discuss this further in section 4.
We shall then consider the consequences of a ``length dependent" velocity of light in some depth.
Special relativity (SR) requires an invariant speed of light in a vacuum, the exact speed is not
specified and so could be larger than we have taken it to be. However, SR does require this speed to
be invariant and we shall see that this alone will cause problems if we do not introduce some finite
length scale which all inertial observers agree upon.

\section{Does the Scharnhorst effect violate Special Relativity?}

\noindent
The standard argument against the violation of SR is that the mirrors break
Lorentz invariance (or more generally Poincaré symmetry, which is the main symmetry group of special
relativity). This implies that the frame has a preferred direction (perpendicular to the mirrors)
and so is no longer inertial. I would counter
by saying that without specifying a length scale the same argument could be used
for most quantum optics experiments where mirrors are used. In fact there would be few
laser related experiments which could be considered in an inertial frame in that case, yet
they never--the--less employ SR.\\

\noindent
Consider the case of a light clock, a pair of mirrors with a light pulse
bouncing between them. This type of clock has been used to
derive effects like time dilation and Lorentz contraction in
undergraduate text book accounts of SR. For time
dilation the clock moves in a direction parallel to the surface of the
mirrors.  Since the Scharnhorst effect predicts that the vacuum
refractive index is equal to unity in that direction then we would expect
the light clock (in a Scharnhorst type arrangement) to give much
the same prediction as in SR. It should be noted however that the
speed of light in the moving frame of the clock has altered due to the
Scharnhorst effect, so predictions are not exactly the same as
with no Scharnhorst effect.
In the Lorentz contraction arrangement, the light
clock is tilted on its side. The direction of motion of the clock is
the same as the light bouncing between the mirrors, (ie. perpendicular
to the mirror surface). It appears that several light
clocks in relative motion
would not calculate the same length scales if the Scharnhorst effect
were operating between the mirrors. Observers in relative motion
would not know what to use for c in the SR velocity addition formula.
The change in the velocity of light, as predicted by the Scharnhorst
effect, is inversely proportional to the fourth power of the
distance between the mirrors, see $\delta c$ below.
The distance measured between the mirrors differs from
one reference frame to another in relative motion due to
Lorentz contraction.
Hence, the measured value of the velocity of light, measured
from one frame to another, must also change and this violates SR
in a very fundamental way.
The whole of SR is based on c being an invariant. Which invariant value
of c do you use in this case? It appears SR does indeed break down if the
Scharnhorst effect were real, even if you cannot measure the
Scharnhorst change in c in practice.\\

Two papers have appeared both stating that measurement of
faster--than--c signals between
mirrors was impossible. The first was by Milonni and Svozil \cite{pwm},
which uses an argument based on the uncertainty relation for velocity
and the uncertainty in time due to switching on a signal.
Their argument can be summarized as follows; A measurement of velocity
involves a distance and time measurement $v=L/t$. The time is limited
by a signal turn on time $\delta t \approx 1/\w$ where $\w$ is the
frequency of the signal. We assume the shortest possible delay. Then
$\delta v = L \delta t/ t^2 \geq c^2 \delta t/L$. Using
$c/\w \approx \lambda$ we obtain $\delta v = c \lambda/L$.
The change in the velocity of light predicted by the Scharnhorst effect
is $\delta c = kc \alpha^2 (\lambda_c /L)^4 $ where
$\lambda_c = \hbar/(m c)= 3.9 \times 10^{-11}$cm is the Compton wavelength,
$\alpha = 1/137$ and $k \approx 10^{-2}$.
Hence the ratio of
\be
\frac{\delta v }{\delta c } \geq \frac{1}{k \alpha^2}
\left( \frac{L}{\lambda_c}\right)^3 = 1.5 \times 10^{6}
\left( \frac{L}{\lambda_c}\right)^3
\ee
where we have used $\lambda \approx \lambda_c$. Thus the measured
uncertainty in velocity, $\delta v$, is much greater than the predicted
change in the signal velocity of light, $\delta c$, by many orders of
magnitude. Thus the Scharnhorst effect is not useful as an experimental
verification that faster--than--c speeds for signals are possible.
Milonni and Svozil also conclude \cite{pwm},
\begin{quote}
`` ... it is clear that the uncertainty in the measured propagation
velocity will always be enormously larger than the correction to c associated
with the Scharnhorst effect. We conclude, therefore, that no measurement
of the faster--than--c velocity of light predicted by the Scharnhorst
effect is possible."
\end{quote}
The second paper by Ben--Menahem \cite{ben}, uses an argument based on the
sharpness of the wavefront.
\begin{quote}
``... in order to observe faster--than--c propagation of the wavefront, it
is necessary to sharpen the falloff of the fields at the wavefront to a length
scale less than $1/m$ [where $m$ is the electron mass].
This feat requires the inclusion in the packet of waves
with $\w > m$, for which eq. (3) [the vacuum refractive index derived by
Scharnhorst for $\w \ll m$] is a bad approximation."
\end{quote}

\noindent
Even if the Scharnhorst effect is real in principle, in practice it is impossible to measure,
which Barton and Scharnhorst agree with themselves. However, it is still objectionable on the grounds
that it would fundamentally violate axioms of QED when no invariant length scale is defined.
It should be stressed that the enhancement of the speed of light predicted by the Scharnhorst effect
is inversely proportional to $L^4$. This variable length is disturbing. Without a fundamental length
scale that all inertial observers agree upon the speed of light becomes an invariant which is in direct
violation of SR.
The Scharnhorst effect may be exactly the kind of thought experiment we need to define an invariant
length scale. This thought experiment would be most noticeable, at nuclear
length scale of a femtometer (fm).
Nuclear length scales are also when true quantum vacuum effects are likely to be observed.
For example, pair production is likely when a charged ion passes close to a nucleus. We discuss
the length scales in the next section.

\section{Discussion and Conclusions}

We have shown in section 3, that although the Scharnhorst effect suggested
that faster--than--c signals were in principle possible, in practice it
would be impossible to detect any such increase in c. We further suggested
that the original calculations of the effect may be oversimplified, and that
the effect implies a violation of SR. We will now consider the effect of the length
scale and validity of the dispersion relation. There is usually a fundamental length scale
defined for a dispersive medium. This length scale is
defined by assuming the wave length of the light under observation must be
much larger than the interatomic spacing of the material. For us to be able to use the dispersion
relations properly for the vacuum what should be the length scale ? Do dispersion relations
apply to the vacuum? Early discussions of this question are by Toll in 1956 \cite{toll}.
Dispersion relations were first used in QED by Gell--Mann, Thirring and Goldberger in
1954 \cite{goldberger}.
Before this we have the paper by Weisskopf 1936 on the electrodynamics of the vacuum, written in
German \cite{weisskopf}.

The dispersion relations, or the real and
imaginary parts of $n(\w)$ where the frequency $\w$ is purely real,
can be written as
\bea
\mbox{Re}\left[ n(\w) \right] &=& 1 + \frac{2 P}{ \pi} \int_0^{\infty}
\frac{ \nu \mbox{Im}\left[ n(\nu) \right] d\nu}{ \nu^2 - \w^2 } \nn \\
&=&  1 + \frac{c P}{ \pi} \int_0^{\infty}
\frac{\alpha_0(\nu) d\nu}{ \nu^2 - \w^2 } \label{kk} \\
\mbox{Im}\left[ n(\w) \right] &=& -\frac{ 2P}{\pi} \int_0^{\infty}
\frac{ \w \mbox{Re}\left[ n(\nu) -1 \right] d\nu}{ \nu^2 -\w^2 }
\eea

\noindent
where $\alpha_0 = 2 n_i ( \w )\w/c$ is the absorption
coefficient, and $n_i$ is the imaginary part of the refractive
index. This is true only when we use a physical model, like the Drude model. The more general results
are written earlier, see equation [\ref{lasteqn}].
The above results are consistent with the findings of the text books
\cite{jack,sak,panof} and also Nussenzveig \cite{nuss}, Toll \cite{toll},
Landau and Lifshitz \cite{stat} and Rauch and Rohrlich \cite{rohr}, all of which assume
a physical model for the material. None of these texts are referring specifically to the vacuum
as the medium.

Historically, the result of Eq.(\ref{kk}) was first derived by
Kronig (1926) \cite{kronig}
and the equivalent result for the dielectric constant  was treated by
Kramers in (1927) \cite{kramer}. Kronig was interested in the physical
model behind the derivation of the refractive index and
how many atoms are required before you can sensibly use the
refractive index idea for a bulk material, he suggested the number 25.
Kramers first employed Cauchy integrals to derive the above
dispersion relations. However, neither paper puts any emphasis on
causality.  This comes in much later by Kronig (1942) \cite{kronig2}.
More recent tutorial accounts of the dispersion relations which
include a discussion on causality
can be found, for example see \cite{peter,wolf}.
Text book accounts \cite{jack,sak,panof},
will mention that $n(\w) \rightarrow 1$
as $\w \rightarrow \infty$ but will only argue based on the physical
model for the refractive index. We have shown in a previous work \cite{heidi1, heidi2} that if
$\left[ n(\w) -1 \right]$
satisfies the dispersion relations then, regardless of how it was
derived, Titchmarsh's theorem \cite{titch} proves that $\left[ n(\w) -1 \right]$
obeys causality and must also therefore have the limiting result
that $n(\w) \rightarrow 1$ for $\w \rightarrow \infty$ by condition (ii) of
Titchmarsh's theorem.\\

As to the question whether or not the dispersion relations are valid for the vacuum. The author would
suggest yes, but that some length scale (of validity) should be imposed.
At micron length scale ( 1 $\mu m$) the increase in the velocity of light
predicted by the Scharnhorst effect is much too small to detect $\Delta c /c = 1.6 \times 10^{-36} $.
At nuclear length scales
of 1 $fm$ would imply a Scharnhorst increase in the velocity of light by $\Delta c /c = 1.6 $.
Elementary particle physics, for example string theory, requires a fundamental length scale,
the Planck length is usually quoted at $ 1.6 \times 10^{-35} m$. This is also much to small to
be measurable any time soon. Perhaps a new length scale defined by the onset of truly quantum vacuum
behavior (pair production or nonlinear photon--photon scattering length scale) is required.
Assuming SR still applies, what is the
invariant speed of light at that scale since to the best of the authors knowledge the speed of light
has not been measured at the nuclear scale?

It is well known that in QED the relativistic causality requirement is imposed by the quantization
procedure. Once we have the Hamiltonian for our quantum system, the field operators must obey
certain equal time commutators for the Hamilton equation of motion to give back the wave equation.
The point like particles must interact only at a point, so if two locations $(x,y)$ are specified
then the equal time commutator or anti-commutator requires $ \delta( x-y)$ , that is $x=y$ for
equal times. This is another way of saying that signals cannot travel faster than light. It is well
summarized in the paper by Gell--Mann, Goldberger and Thirring 1954 \cite{goldberger}:

\begin{quote}
``The causality requirement in the present paper is as follows; The quantum mechanical formulation
of the demand that waves do not propagate faster than the velocity of light is, as is well known,
the condition that the measurements of two observable quantities should not interfere if the two
points of measurement are space like to each other."
\end{quote}

Finally, we point out that invariant length scales already exist, the so called Doubly special relativity
makes use of an invariant Planck length scale and this leads directly to nonlinear electromagnetism.
Special relativity at low energies can be preserved in this process, \cite{smolin1,smolin2,camelia}.
Perhaps a larger invariant length scale of order $ 1 fm$ might be more appropriate. This would seem
to suggest the onset of pair production (via ion nuclear interaction) and also it appears
to be suggested by the Scharnhorst effect (if the dispersion relations are valid)
as a scale where modifications to the speed of light might be noticeable.

\section*{Acknowledgement}

\noindent
This work was prepared specifically for the conference proceedings of the Coherent
Control of the Fundamental Processes in Optics and X-ray optics (CCFP'06), held
in Nizhny Novgorod June 29th -July 3rd 2006. The author wishes to thank the organizers
of the conference, and specifically Olga Kocharovskaya and Mikhail Tokman.\\


\end{document}